\documentclass[aps,twocolumn,a4paper,pra,superscriptaddress,floatfix]{revtex4}%
\usepackage{amssymb}
\usepackage{amsmath}
\usepackage{amsfonts}
\usepackage{graphicx}
\usepackage{graphics}

\def\P{\mathrm{P}}    
\def\S{\mathrm{S}}    
\def\PS{\mathrm{PS}}  
\def\TE{\mathrm{TE}}  
\def\TM{\mathrm{TM}}  
\def\bk{\mathbf{k}}
\def\dd{\mathrm{d}}
\def\PFA{\mathrm{PFA}}   
 
\def\E{\mathrm{E}}
\def\M{\mathrm{M}}

\def\txi{\widetilde \xi}
\def\scalar{\mathrm{sc}}   
\def\em{\mathrm{em}}   

\begin{document}

\title{Casimir energy between a plane and a sphere in electromagnetic vacuum}

\author{Paulo A. Maia Neto}
\affiliation{Instituto de F\'{\i}sica, UFRJ, 
CP 68528,   Rio de Janeiro,  RJ, 21941-972, Brazil}

\author{Astrid Lambrecht}
\author{Serge Reynaud}
\affiliation{Laboratoire Kastler Brossel,
CNRS, ENS, Universit\'e Pierre et Marie Curie case 74,
Campus Jussieu, F-75252 Paris Cedex 05, France}

\date{\today}

\begin{abstract}
The Casimir energy is computed in the geometry of interest for the most precise 
experiments, a plane and a sphere in electromagnetic vacuum. 
The scattering formula is developed on adapted plane-waves and multipole basis,
leading to an expression valid for arbitrary relative values of 
the sphere radius and inter-plate distance.
In the limiting case of perfect reflection, the electromagnetic result
is found to depart from the commonly used proximity-force approximation (PFA) 
significantly more rapidly than expected from scalar computations.
\end{abstract}


\maketitle

The Casimir force is a mechanical effect of quantum vacuum fluctuations \cite{Casimir} 
with a large impact in micro- and nanotechnology \cite{Roukes01,Chan01}. 
Experimental advances allowing accurate measurements of small surface forces 
at micrometric distances have led to a number of precise Casimir force measurements 
in the last 10 years (see reviews in \cite{Lamoreaux,Bordag,LambrechtNJP06}).
These measurements and their comparison with Quantum ElectroDynamics (QED) predictions 
have become a powerful tool for searching for the presence of the new hypothetical 
forces which are predicted by the models aiming at unifying gravity with quantum
theory \cite{OnofrioNJP06,DeccaPRD07}.
As the force varies substantially with the experimental conditions, 
accurate theoretical computations of realistic systems 
are needed for these comparisons to be reliable and fruitful.
The influences of imperfect reflection \cite{LambrechtNJP06} 
and non null temperature \cite{BrevikNJP06}, which have 
been studied extensively, are not discussed further here. 

Instead we will focus our attention on the rich connection 
between Casimir effect and geometry \cite{Balian}.
As the most precise measurements performed to date involve a plane 
and a sphere, the effect of geometry is important for the purpose of 
theory-experiment comparison.
It is usually calculated through the Proximity Force Approximation 
(PFA) \cite{Derjaguin68} which amounts to average the plane-plane
expression over the distribution of interplate distances.
This approximation can only be valid \cite{Schaden00,Jaffe04}
at the limit where the sphere $R$ is much larger than the 
inter-plate separation $L$. Even in this limit,
it is still worth specifying its accuracy in order to 
master the quality of theory-experiment comparison \cite{Krause07}.

\textit{The purpose of this work -}
A number of results going beyond the PFA have been obtained 
\cite{Gies03,JaffePRA05,Bulgac06,Bordag06,Emig06,Dalvit06,Emig07,%
Milton07,Rodriguez07,WirzbaQFEXT07,BordagQFEXT07,Emig08}.
It is only very recently that results were obtained for the 
case of direct relevance for the most precise experiments,
namely the configuration of a plane and a sphere in
electromagnetic vacuum \cite{Emig08}.
It is the purpose of the present work to compute 
the Casimir energy in this configuration and draw
consequences for theory-experiment comparison.

\begin{figure}[t]
\centering
\includegraphics[width=3cm]{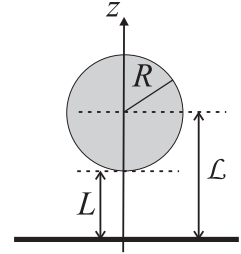}
\caption{
Plane-sphere geometry. }
\end{figure}

We study the case of a sphere (radius $R$) on top of a plane plate (see Fig.~1).
We denote $\cal L$ the center-to-plate distance and $L$ the inter-plate distance.
The plane-sphere Casimir energy ${\cal E}_\PS$ will be written
in terms of a ratio to the PFA formula (here written for perfect reflection)
\begin{eqnarray}
\label{defrho}
 \rho_\PS \equiv \frac{{\cal E}_\PS} {{\cal E}_\PS^\PFA}  
\;,\quad 
{\cal E}_\PS^\PFA  = -\frac{\hbar c\pi ^3 R}{720 L^2} &&
\end{eqnarray} 
Computations performed with scalar models \cite{WirzbaQFEXT07,BordagQFEXT07} 
can be used to guess the expression of the ratio $ \rho_\PS$ for small values of $L/R$.
Considering that the electromagnetic case is given by the sum of results obtained
with Dirichlet and Neumann boundary conditions, one thus gets
\begin{eqnarray}
\label{scalar}
\rho_\PS = 1-\frac{\nu_\scalar L}{R} + O\left(\frac{L^2}{R^2} \right)
\,,\quad \nu_\scalar = \frac{5}{\pi^2}-\frac{1}{3}  &&
\end{eqnarray} 
Bordag and Nikolaev \cite{BordagQFEXT07} remark that the electromagnetic result 
does not necessarily meet the scalar expression (\ref{scalar}), 
but expect their magnitudes to be close to each other.
The main result of the present work will be that the beyond-PFA corrections,
as measured by the factor $\nu$, is in fact
significantly larger in the electromagnetic case than in the scalar estimate.

\textit{General scattering formula -}
Our starting point is the general scattering formula describing the Casimir energy 
between two scatterers in electromagnetic vacuum \cite{LambrechtNJP06}. 
This formula has been used for dealing with rough or corrugated metallic mirrors
in the limiting case where the roughness or corrugation was treated as a
small perturbation \cite{MaiaNeto05,Rodrigues06,ReynaudJPA08}. The PFA was thus found to be valid 
when the roughness or corrugation wavelength was larger than the other length scales. 
Here, we apply it to the geometry of Fig.~1 where it is read
\begin{eqnarray}
\label{depart}
&&{\cal E}_\PS=\hbar\,\int_0^{\infty} \frac{d\xi}{2\pi}\log{\rm det}\, {\cal D} \\
&&{\cal D} =1-{\cal R}_\S\,e^{-{\cal K} {\cal L}}\,{\cal R}_\P\,e^{-{\cal K} {\cal L}}
\nonumber
\end{eqnarray}
${\cal R}_\S$ and ${\cal R}_\P$ are reflection operators on the spherical and plane scatterers 
while $e^{-{\cal K} {\cal L}}$ represents one-way propagation along the $z$-axis between reference points 
sitting respectively at the centre of the sphere and on the plane.
In the following, we write (\ref{depart}) in a more explicit manner by introducing 
adapted mode basis.

The plane-wave basis  $|\bk,\phi,p \rangle_{\xi}$ ($\bk $ the transverse
wavevector, $p=\TE,\TM$ the polarization,
$\phi=\pm 1$ the upwards/downwards propagation direction and 
$\xi$ the imaginary frequency) is well adapted to the description
of free propagation and reflection on the plane: 
the operator $e^{-{\cal K L}}$ is diagonal in this basis (matrix elements 
$e^{-\kappa {\cal L}}$ with $\kappa = \sqrt{\xi^2/c^2+k^2}$);
reflection on the plane also preserve all plane wave quantum numbers but $\phi$ 
(the non zero elements of the matrix ${\cal R}_\P$ 
are given by the Fresnel specular reflection amplitudes $r_p$). 

We also use the multipole basis $|\ell m P\rangle_{\xi}$ 
for the subspace corresponding to a given $\xi$; 
$\ell(\ell+1)$ and $m$ are the usual angular momentum eigenvalues
(with $\ell=1,2,...$, $m=-\ell,...,\ell$)
and $P=\E,\M$ denote electric and magnetic multipoles. 
By rotational symmetry around the $z$-axis, $\cal D$ commutes with $J_z$. 
Hence $\cal D$ is block diagonal, and each block ${\cal D}^{(m)}$ (corresponding to a given $m$) 
yields an independent contribution to the Casimir energy.
The elements of this block ${\cal D}^{(m)}$  are given by
\begin{eqnarray}
\label{plane-waves}
\nonumber
{\cal D}^{(m)}_{1,2} & = &\delta_{1,2} - 
\int\frac{d^2\bk}{(2\pi)^2}\sum_{p=\TE,\TM}
\langle \ell_1 m P_1 | {\cal R}_S |\bk,+,p \rangle\\
&& \times  r_p(\bk)e^{-2\kappa {\cal L}} \,\langle \bk,-,p|\ell_2 m P_2\rangle
\end{eqnarray}
When read from right to left, this expression has the following interpretation: 
a multipole wave $(\ell_2 m  P_2)$ is first decomposed into plane waves which 
propagate towards the (plane) plate where it is reflected;
it then propagates back to the sphere, and is finally scattered 
into a new multipole wave $(\ell_1 m P_1).$

Reflection on the sphere can be written in terms of Mie coefficients $a_{\ell}(i \txi)$ 
and $b_{\ell}(i\txi)$, corresponding to electric and magnetic multipoles 
respectively \cite{Bohren} and of the finite rotation matrix \cite{Edmonds} elements 
$d^{\ell}_{m,1}(\theta)$. 
The former depend on the reduced parameter $\txi= \xi R/c$
and the latter on the angle $\theta$ such that
$\cos\theta=c\kappa/\xi \ge 1,$ $\sin\theta=-ick/\xi$,
both evaluated for the imaginary frequency $\xi$.
As soon as the Mie coefficients are computed, as the
specular reflection coefficients \cite{EPJD}, from the optical response 
of the sphere and plane, the scattering 
formula~(\ref{plane-waves}) allows one to obtain the Casimir energy 
for arbitrary separation distances and arbitrary isotropic materials.

\textit{Limit of perfect reflectors -}
In the present work, we focus the attention on the case of
perfectly-reflecting plane and sphere, which can be directly
compared to already available scalar results.
Precisely, we consider metallic materials described by the plasma 
model \cite{EPJD}, with a plasma wavelength $\lambda_\P$.
We then obtain a proper definition of perfect reflectors as the limit 
where $\lambda_\P$ is smaller than the length scales $R$ and $L$.
The Mie coefficients can thus be written in terms of the modified 
Bessel functions (defined as in \cite{Abramowicz})
\begin{eqnarray}
\label{Mie}
&& a_{\ell}(i\txi)= \frac{\pi}{2}(-1)^{\ell+1}
\frac{  \ell I_{\ell+1/2}(\txi)-\txi  I_{\ell-1/2}(\txi)}
{ \ell K_{\ell+1/2}(\txi)+\txi  K_{\ell-1/2}(\txi)}
\nonumber \\
&& b_{\ell}(i\txi)= \frac{\pi}{2}(-1)^{\ell+1}
\frac{  I_{\ell+1/2}(\txi)}
{  K_{\ell+1/2}(\txi)}. 
\end{eqnarray}
Meanwhile, specular reflection on the plane is described by $r_{\TM}=-r_{\TE}=1$.
In this case, we are able to sum up the TE and TM contributions 
in (\ref{plane-waves}) to get analytical expressions for the matrix elements
\begin{eqnarray}
\label{EEMM}
& {\cal D}^{(m)}_{\ell_1 \E,\ell_2 \E} = \delta_{\ell_1\ell_2}
+\frac{1}{2}\sqrt{(2\ell_1+1)(2\ell_2+1)} a_{\ell_1}
{\cal F}^{(+)}_{\ell_1,\ell_2,m} \nonumber \\
& {\cal D}^{(m)}_{\ell_1 \M,\ell_2 \M} = \delta_{\ell_1\ell_2}
-\frac{1}{2}\sqrt{(2\ell_1+1)(2\ell_2+1)} b_{\ell_1}
{\cal F}^{(+)}_{\ell_1,\ell_2,m} \nonumber \\
& {\cal D}^{(m)}_{\ell_1 \E,\ell_2 \M} = 
\frac{i}{2}\sqrt{(2\ell_1+1)(2\ell_2+1)} a_{\ell_1}
{\cal F}^{(-)}_{\ell_1,\ell_2,m} \nonumber \\
& {\cal D}^{(m)}_{\ell_1 \M,\ell_2 \E} = 
\frac{i}{2}\sqrt{(2\ell_1+1)(2\ell_2+1)} b_{\ell_1}
{\cal F}^{(-)}_{\ell_1,\ell_2,m} 
\end{eqnarray}
The overlap integrals ${\cal F}$ 
\begin{eqnarray}
\label{F}
&{\cal F}^{(\pm)}_{\ell_1,\ell_2,m} = (-)^{\ell_2+m} \int_1^{\infty} \dd\cos\theta
 \;e^{-2\xi {\cal L} \cos\theta /c}\\
\nonumber
& \times \Biggl[ d^{\ell_1}_{m,1}(\theta) d^{\ell_2}_{m,1}(\theta) \pm (-)^{\ell_1-\ell_2}
d^{\ell_1}_{m,1}(\pi-\theta) d^{\ell_2}_{m,1}(\pi-\theta)
 \Biggr] 
\end{eqnarray}
contain a factor $\exp(-2\xi{\cal L}/c)$ which play a key
role in the foregoing discussion of distance dependence of ${\cal E}_\PS $. 

When the distance is larger than the radius, or equivalently when the sphere is small
$R\ll {\cal L}$, the Mie coefficients are needed for small `size parameters' $\txi \ll 1.$
The dominant contribution thus comes from $\ell =1$ with
$a_1({\txi})\approx -2b_1({\txi}) \approx -2{\txi}^3/3$.
The Casimir energy is then obtained from 
${\cal D}^{(0)}$ and ${\cal D}^{(1)},$ approximated by $2\times 2$ matrices ($P=\E,\M$). 
The product of the nondiagonal elements of these matrices is found to be negligible, 
and the following result obtained at lowest order in $R/{\cal L}$ 
\begin{eqnarray}
\label{smallsphere}
{\cal E}_\PS = -\frac{9\,\hbar c }{16\pi} \frac{R^3}{{\cal L}^4} 
\quad,\quad \left[\,\lambda_\P\ll R \ll {\cal L}\,\right] &&
\end{eqnarray}

Since the magnetic dipole contribution is one-half of the electric dipole one, 
the small-sphere limit (\ref{smallsphere}) is $50\%$ larger than the 
Rayleigh limit, calculated by accounting only for the electric contribution.
Note that the latter can also be recovered from the general scattering formulas
(\ref{depart}-\ref{plane-waves}), but not when using the Mie coefficients 
for the perfectly-reflecting sphere (\ref{Mie}). 
As a matter of fact, perfect reflection corresponds to $\lambda_\P$ being the
smallest length scale whereas the Rayleigh limit corresponds to 
$R\ll\lambda_P\ll {\cal L}$, with a negligible magnetic contribution.

As the sphere radius increases, higher values of $\ell$ and ${\txi}$ 
become increasingly important. 
The Mie coefficients grow as $\exp(2{\txi})$ for ${\txi}\gg 1.$
When multiplied by the overlap integrals in (\ref{EEMM}), they 
produce a factor $\sim \exp(-2\xi L/c)$.
Using the `localization principle' \cite{Moyses}, 
 we may estimate the values of $\ell$ contributing appreciably to
the Casimir energy for a given value of $L/R.$  A given angular momentum $\ell$ 
corresponds semiclassically to an impact parameter $B=c\ell/\xi.$  If $B>R,$ 
its contribution is negligible since it corresponds to `rays' that do not hit the sphere. 
With $\xi \sim c/L,$ we then expect that $\ell \gg R/L$ provide negligible contributions. 
Numerical confirmations of this fact are presented below.

\textit{Result of the numerical evaluation -}
We have numerically evaluated the ratio $\rho_\PS$ of the plane-sphere Casimir energy 
normalized to the PFA expectation (see eq.\ref{defrho})
\begin{eqnarray}
\rho_\PS = -\frac{360}{\pi^4} \frac{L ^2 }{R ^2 } 
\int_0^{\infty} d\tilde \xi 
 \sum_{m=-\infty}^\infty \log \det {\cal D}^{(m)}  &&
\end{eqnarray} 
{}For perfectly-reflecting plates,
this ratio is a function of the single parameter $L/R$,
that we plot on Fig.~2.
Several important features can be noticed on this plot. 
{}First, the small-sphere limit (blue curve) provides a good approximation 
of the exact result (black curve) from large values of $L/R$ 
down to $L/R \sim 5$. 
In the short distance limit $L/R\rightarrow0$, the exact result
goes to the PFA expectation $\rho_\PS=1$ whereas 
the small-sphere approximation leads to a strong underestimation. 
Note that we always have $\rho_\PS<1$, which means that 
the PFA systematically overestimates the Casimir energy. 

\begin{figure}[t]
\centering
\includegraphics[width=8cm]{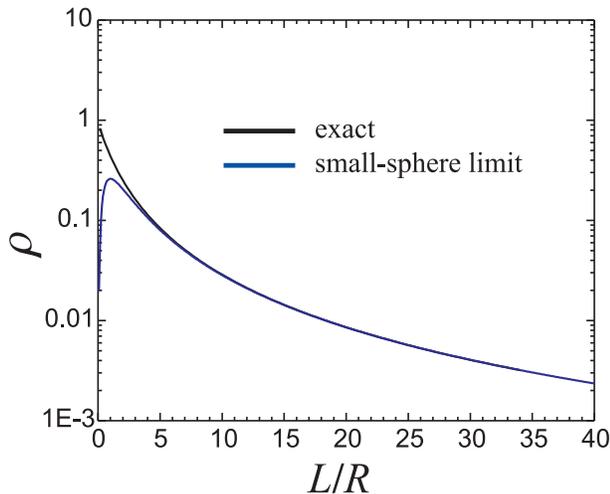}
\caption{
Ratio $\rho_\PS$ showing the deviation from PFA as a function of $L/R$. 
The black and blue curves represent respectively the `exact' result
and the small-sphere limit (both for perfectly-reflecting plates). 
}
\end{figure}

In order to discuss the most precise experiments, we zoom out 
on the interval $L/R<2$ on Fig.~3, 
and devote a more detailed analysis to it. 
We first address the increased difficulty of the numerical
evaluation near the PFA (see \cite{EmigQFEXT07} for a similar 
discussion in the geometry of two equal spheres). 
As $L/R$ is decreased towards the PFA limit, 
larger and larger values of $\ell$ are needed. 
The localization principle requires $\ell>\alpha\, R/L$
where $\alpha$ is a numerical value that we have found to be 
approximately 4.
The curves on Fig.~3 are numerical evaluations of $\rho_\PS$ respectively
for maximum angular momentum $\ell_{\rm max}=10,15,27$. 
The vertical dashed lines indicate the values $L/R= 0.4$ and 
$L/R= 0.27$ where the curves evaluated for $\ell_{\rm max}=10$ and 
15 depart from the better calculation with $\ell_{\rm max}=27$.
Using the same argument, we predict that the curve computed with $\ell_{\rm max}=27$
should be accurate down to $L/R=0.16$ (also indicated by a dashed vertical line). 
It is worth doing computations with larger $\ell_{\rm max}$
for increasing this range of validity \cite{NumericalNote}.
But it also clear that these computations are expected
to produce points close to the dots drawn by joining our numerical results 
to the PFA limit $\rho_\PS(L=0)=1$.

\begin{figure}[t]
\centering
\includegraphics[width=8cm]{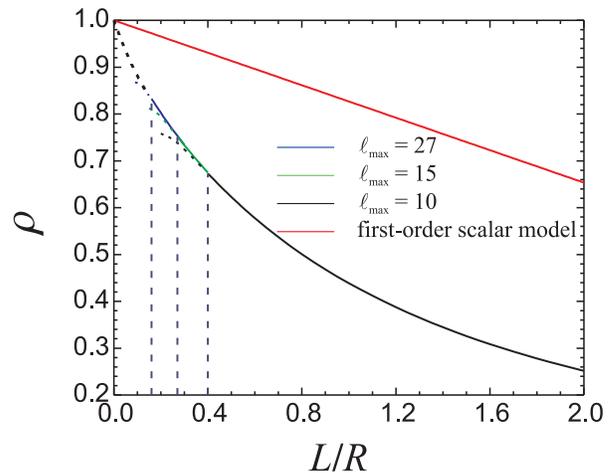}
\caption{
Zoom of Fig.~2 for $L/R<2$. 
The green, blue and black curves are numerical evaluations of $\rho_\PS$ with
maximum angular momentum $\ell_{\rm max}=10,15,27$. 
The vertical dashed lines indicate the limit of validity of these evaluations.
The red curve shows the scalar result (eq.\ref{scalar}) at first order in $L/R$.
The dots correspond to a best-fit quadratic function of $L/R$ joining
our numerical results to the PFA limit.}
\end{figure}

\textit{Comparison with existing results -}
In order to compare the present electromagnetic results 
with those obtained with scalar computations 
\cite{WirzbaQFEXT07,BordagQFEXT07}, we have drawn
the scalar prediction (\ref{scalar}), truncated at first order in $L/R$, 
as the red curve on Fig.~3.
Clearly, it does not fit the result we have obtained
for the electromagnetic case.
Precisely, the beyond-PFA correction appears to be 
several times larger in the electromagnetic case
than in the scalar one.
In order to make this remark more quantitative, we obtain a best-fit
quadratic function $\rho_\PS=1-\nu_\em L/R+\nu_2 L^2/R^2$ joining the PFA limit
$\rho_\PS(L=0)=1$ to our numerical results. We thus get an
estimate of the parameter $\nu_\em$ defined as in (\ref{scalar}) 
but for the electromagnetic computation
\begin{eqnarray}
\nu_\em \sim 1.4 \sim 8 \times \nu_\scalar &&
\end{eqnarray} 
This means that the Casimir energy between a plane and a sphere 
in electromagnetic vacuum departs from the PFA expectation significantly
more rapidly than expected from scalar computations. 
Our result agrees with that obtained in \cite{Emig08}, and it has been 
obtained through an independent calculation based on a different approach 
(\textit{i.e.} the scattering approach of \cite{LambrechtNJP06}).
It constitutes a very important hint to be included in the 
discussion of the quality assessment of theory-experiment 
comparisons in the plane-sphere geometry \cite{Krause07}.

This result has been obtained in the limiting case of perfect reflectors,
for which some existing results were already available.
Clearly, extra work is needed before definitive consequences 
can be drawn for theory-experiment comparisons.
As a matter of fact, the most precise experiments to date
are performed at inter-plate distances not significantly larger
than the plasma wavelength, and this entails that the effect of 
metallic response of the materials plays a non negligible role.
However, the results already in our hands are a clear warning
that the effect of plane-sphere geometry has to be treated with 
the greatest care when comparing the measured plane-sphere Casimir energy
with QED theoretical predictions.

\acknowledgements 
The authors thank M.T. Jaekel, C. Genet, I. Cavero-Pelaez, D.A.R. Dalvit, 
D. Delande, B. Gremaud and V. Nesvizhevsky for stimulating discussions.
P.A.M.N. thanks  CNPq, CAPES, Institutos do Mil\^enio de Informa\c c\~ao 
Qu\^antica e Nanoci\^encias for financial support and ENS for a visiting
professor position. 
A.L. acknowledges partial financial support 
by the European Contract STRP 12142 NANOCASE.

\newcommand{\REVIEW}[4]{\textrm{#1} \textbf{#2}, #3, (#4)}
\newcommand{\Review}[1]{\textrm{#1}}
\newcommand{\Volume}[1]{\textbf{#1}}
\newcommand{\Book}[1]{\textit{#1}}
\newcommand{\Eprint}[1]{\textsf{#1}}
\def\etal{\textit{et al}}


\begin{thebibliography}{99}

\bibitem{Casimir}  H.B.G. Casimir, 
\Review{Proc. K. Ned. Akad. Wet.} \Volume{51}, 793 (1948).

\bibitem{Roukes01} E. Buks, M.L. Roukes, 
\Review{Phys. Rev.} \Volume{B63}, 033402 (2001).

\bibitem{Chan01} H.B. Chan, V.A. Aksyuk, R.N. Kleiman \etal, 
\Review{Science} \Volume{291}, 1941 (2001); 
\Review{Phys. Rev. Lett.} \Volume{87}, 211801 (2001).

\bibitem{Lamoreaux} S. K. Lamoreaux
\Review{Am. J. Phys.} \Volume{67}, 850 (1999)

\bibitem{Bordag} M. Bordag, U. Mohideen and V.M. Mostepanenko,
\Review{Phys. Rep.} \Volume{353}, 1 (2001).

\bibitem{LambrechtNJP06}  A. Lambrecht, P.A. Maia Neto and S. Reynaud, 
\Review{New J. Phys.} \textbf{8}, 243 (2006).

\bibitem{OnofrioNJP06}  R. Onofrio, 
\Review{New J. Phys.} \textbf{8}, 237 (2006). 

\bibitem{DeccaPRD07}  R. Decca, D. Lopez, E. Fischbach \etal,
\Review{Phys. Rev.} \textbf{D75}, 077101 (2007). 

\bibitem{BrevikNJP06} I. Brevik, S.A. Ellingsen and K. Milton,
\Review{New J. Phys.} \textbf{8}, 236 (2006). 

\bibitem{Balian} R. Balian and B. Duplantier, 
\Review{Ann. Phys. NY} \textbf{104}, 300 (1977); 
\textbf{112}, 165 (1978); in 
{\it 15th SIGRAV Conference on General Relativity and Gravitation}, 
[\Eprint{arXiv:quant-ph/0408124}].

\bibitem{Derjaguin68}  B.V. Deriagin, I.I. Abrikosova and E.M. Lifshitz,
\Review{Quart. Rev.} \Volume{10}, 295  (1968).

\bibitem{Schaden00} M. Schaden and L. Spruch,
\Review{Phys. Rev. A} \textbf{58}, 935 (1998);
\Review{Phys. Rev. Lett.} \textbf{84}, 459 (2000).

\bibitem{Jaffe04}   R.L. Jaffe and A. Scardicchio,
\Review{Phys. Rev. Lett.} \textbf{92}, 070402 (2004).

\bibitem{Krause07} D.E. Krause, R.S. Decca, D. Lopez and E. Fischbach, 
\textrm{Phys. Rev. Lett.} \textbf{98}, 050403 (2007).

\bibitem{Gies03} K. Langfeld, L. Moyaerts and H. Gies, 
\Review{J. High En. Phys.} \Volume{0306}, 018 (2003);
H. Gies and K. Klingm\"uller, 
\Review{Phys. Rev. Lett.} {\bf 96}, 220401 (2006).

\bibitem{JaffePRA05} O. Schr\"oder, A. Sardicchio and R.L. Jaffe, 
\Review{Phys. Rev.} \Volume{A72}, 012105 (2005).

\bibitem{Bulgac06} A. Bulgac, P. Magierski and A. Wirzba, 
\Review{Phys. Rev.} {\bf D73}, 025007 (2006).   

\bibitem{Bordag06} M. Bordag, 
\Review{Phys. Rev.} {\bf D73}, 125018 (2006).

\bibitem{Emig06}  T. Emig, R.L. Jaffe, M. Kardar and A. Scardicchio, 
\Review{Phys. Rev. Lett.} {\bf 96}, 080403 (2006).

\bibitem{Dalvit06}  D.A.R. Dalvit, F.C. Lombardo, F.D. Mazzitelli 
and R. Onofrio, \Review{Phys. Rev.} A {\bf 74}, 020101 (2006);
F.D. Mazzitelli, D.A.R. Dalvit and F.C. Lombardo,  
\Review{New J. Phys.} \textbf{8}, 240 (2006). 

\bibitem{Emig07}   T. Emig, N. Graham, R.L. Jaffe and M. Kardar,
\Review{Phys. Rev. Lett} {\bf 99}, 170403 (2007).

\bibitem{Milton07}   K.A. Milton and J. Wagner,
\Eprint{arXiv:0711.0774}; \Eprint{arXiv:0712.3811}.

\bibitem{Rodriguez07} A. Rodriguez, M. Ibanescu, D. Iannuzzi \etal, 
\textrm{Phys. Rev. Lett.} \textbf{99}, 080401 (2007).

\bibitem{WirzbaQFEXT07} A. Wirzba, \Review{J. Phys.} \Volume{A41} 164003 (2008).

\bibitem{BordagQFEXT07} M. Bordag and V. Nikolaev, \Review{J. Phys.} \Volume{A41} 164002 (2008).

\bibitem{Emig08}   T. Emig, \Review{J. Stat. Mech.}, P04007 (2008).

\bibitem{MaiaNeto05} P.A. Maia Neto, A. Lambrecht and S. Reynaud,
\Review{Europhys. Lett.} \Volume{69}, 924 (2005); 
\Review{Phys. Rev.} \Volume{A72}, 012115 (2005).

\bibitem{Rodrigues06} R.B. Rodrigues, P.A. Maia Neto, A. Lambrecht and S. Reynaud,
\Review{Phys. Rev. Lett.} \Volume{96}, 100402 (2006);
\Review{Europhys. Lett.} \Volume{76}, 822 (2006); 
\Review{Phys. Rev.} \Volume{A75}, 062108 (2007).

\bibitem{ReynaudJPA08}  S. Reynaud, P.A. Maia Neto and A. Lambrecht, 
\Review{J. Phys.} \Volume{A41} 164004 (2008).

\bibitem{Bohren} C. F. Bohren and D. R. Huffman, 
{\it Absorption and Scattering of Light by Small Particles}
(Wiley, New York, 1983) ch. 4. 

\bibitem{Edmonds} A. R. Edmonds, 
{\it Angular Momentum in Quantum Mechanics},
(Princeton University Press, 1957).

\bibitem{EPJD}  A. Lambrecht and S. Reynaud, 
\Review{Eur. Phys. J.} \Volume{D8}, 309 (2000).

\bibitem{Abramowicz} M. Abramowitz and I. Stegun,  
{\it Handbook of Mathematical Functions},
(Dover, New York, 1972).

{\it Principles of Optics}, Pergamon, Oxford, 1980, sec. 13.5. 

\bibitem{Moyses}  H. M. Nussenzveig  
{\it Diffraction Effects in Semiclassical Scattering} 
(Cambridge University Press, 1992).

\bibitem{EmigQFEXT07} T. Emig and R.L. Jaffe, \Review{J. Phys.} \Volume{A41} 164001 (2008).

\bibitem{NumericalNote} The main ingredients of our FORTRAN code include a Kronrod-Patterson 
Gaussian-type quadrature method for numerical integration, LU decomposition for the 
determinant evaluation, upwards recurrence for $d^{\ell}_{m,1}$ and $K_{\ell+1/2}({\txi})$ 
and downwards recurrence for $I_{\ell+1/2}({\txi})$. 
Improvements that would allow for computation with larger values of 
$\ell_{\rm max}$ are currently under way.

\end{thebibliography}
\end{document}